\documentclass[letterpaper, 12pt]{ewshm}

\usepackage[dvips,final]{graphicx}

\usepackage[table]{xcolor}
\usepackage{makecell}
\usepackage{enumerate}
\usepackage{color}
\usepackage{caption}
\usepackage{subcaption}
\usepackage{amsfonts}
\usepackage{amsmath}
\usepackage{amssymb}
\usepackage{times}
\usepackage{cite}
\usepackage{hhline}
\usepackage{compactbib}
\usepackage{graphicx}        
\usepackage{amsmath,amsfonts,amssymb}
\usepackage{geometry}
\usepackage[labelsep=period]{caption}
\usepackage{booktabs}
\usepackage{tabularx}
\usepackage{array}

\definecolor{halfgreen}{RGB}{0,128,0}
\definecolor{ahsred}{RGB}{192,0,0}

\renewcommand{\thefigure}{\arabic{figure}}
\renewcommand{\thetable}{\Roman{table}}

\newcommand{\beq}{\begin{equation}}
\newcommand{\eeq}{\end{equation}}
\newcommand{\bgqar}{\begin{eqnarray}}
\newcommand{\enqar}{\end{eqnarray}}
\newcommand{\bgqarn}{\begin{eqnarray*}}
\newcommand{\enqarn}{\end{eqnarray*}}
\newcommand{\bgary}{\begin{array}}
\newcommand{\enary}{\end{array}}

\long\def\symbolfootnote[#1]#2{\begingroup%
\def\thefootnote{\fnsymbol{footnote}}\footnote[#1]{#2}\endgroup}

\makeatletter
\renewcommand\@biblabel[1]{#1.}
\makeatother

\begin{document}


\vspace*{4.4cm}

\noindent Title: \textbf{Comparative Evaluation of Neural Network Architectures for Generalizable Human Spatial Preference Prediction in Unseen Built Environments}

\vspace{1.6cm}

\noindent
$
\begin{array}{ll}
\text{Authors}:  
& \text{Maral Doctorarastoo} \\ 
& \text{Katherine A. Flanigan} \\
& \text{Mario Bergés} \\
& \text{Christopher McComb} 
\end{array}
$

\newpage


\vspace*{60mm}

\noindent \uppercase{\textbf{ABSTRACT}} \vspace{12pt} 

The capacity to predict human spatial preferences within built environments is instrumental for developing 
Cyber-Physical-Social Infrastructure Systems (CPSIS). A significant challenge in this domain is the generalizability of preference models, particularly their efficacy in predicting preferences within environmental configurations not encountered during training. While deep learning models have shown promise in learning complex spatial and contextual dependencies, it remains unclear which neural network architectures are most effective at generalizing to unseen layouts.  To address this, we conduct a comparative study of Graph Neural Networks, Convolutional Neural Networks, and standard feedforward Neural Networks using synthetic data generated from a simplified and synthetic pocket park environment. Beginning with this illustrative case study, allows for controlled analysis of each model's ability to transfer learned preference patterns to unseen spatial scenarios. The models are evaluated based on their capacity to predict preferences influenced by heterogeneous physical, environmental, and social features. Generalizability score is calculated using the area under the precision-recall curve for the seen and unseen layouts. This generalizability score is appropriate for imbalanced data, providing insights into the suitability of each neural network architecture for preference-aware human behavior modeling in 
unseen built environments.



\symbolfootnote[0]{\hspace*{-7mm} Maral Doctorarastoo\textsuperscript{1}, Katherine A. Flanigan, PhD\textsuperscript{2} (Corresponding author), Mario Berg\'es, PhD\textsuperscript{3} (Mario Berg\'es holds concurrent appointments at Carnegie Mellon University (CMU) and as an Amazon Scholar. This manuscript describes work at CMU and is not associated with Amazon.), Christopher McComb, PhD\textsuperscript{4}. Email: \{mdoctora\textsuperscript{1}, kflaniga\textsuperscript{2}, mberges\textsuperscript{3},
ccm\textsuperscript{4}\}@andrew.cmu.edu. Department of Civil and Environmental Engineering, Carnegie Mellon University, Pittsburgh, PA, USA.}


\vspace{24pt} 
\noindent \uppercase{\textbf{INTRODUCTION}}  
\vspace{12pt} 

The design and operation of Cyber-Physical-Social Infrastructure Systems (CPSIS) hinge on the accurate modeling of human spatial behavior and underlying preferences because these systems must adapt the built environment to human needs, support dynamic decision making, and function effectively in real-world environments where human presence, movement, and choices are variable and central to system performance \cite{Doctorarastoo2023CPSISFramework, doctorarastoo2024preference}. By capturing these behavioral patterns, such models support the creation of environments that respond to users, ensuring that social objectives---such as comfort, productivity, and social interaction---are prioritized alongside, and not overshadowed by, economic goals like energy and resource efficiency. While various computational approaches exist to model human spatio-temporal behavior, a persistent challenge is the development of models that exhibit scenario-based generalizability, i.e., the capacity to predict human spatial behavior and preferences in  environmental configurations not seen during model training \cite{doctorarastooModeling, qiao2019scenario}. This aligns with the real-world operation and design problem faced in CPSIS in which actuation needs to be evaluated before implementation.  

Many contemporary preference models struggle to generalize effectively beyond their specific training datasets. The underlying mapping we seek to capture---from high-dimensional environmental stimuli to a probability distribution over individual choices---is a highly nonlinear function that depends simultaneously on physiological, psychological, spatial, environmental, temporal, and social cues \cite{Doctorarastoo2024GNNPreference, Lim_etal_2018, LosonczyMarshall_2013}. 
In principle, this mapping could be approximated by universal function approximators such as deep neural networks, which are capable of representing highly nonlinear relationships. Simpler models, such as linear regressors, may offer interpretability but often lack the capacity to capture the complex, multimodal dependencies inherent in spatial preference data. 
Studies have shown that environmental characteristics like visibility significantly influence seat selection in libraries \cite{Lim_etal_2018}, while factors such as performance goals, social needs, and even cultural background impact choices in classroom settings \cite{LosonczyMarshall_2013, Lu_etal_2024}. 
Deep data-driven machine learning models have demonstrated considerable potential in capturing these complex interactions; however, they often risk overfitting to the training scenarios, which limits their performance when applied to new situations \cite{tenzer2023geospatial}.

Neural network architectures offer different inductive biases for learning from influential features. Graph Neural Networks (GNNs), have recently gained attention for their proficiency in modeling relational data \cite{honarvar2024geometric, taghizadeh2025interpretableGNN}, a characteristic that makes them well-suited for representing the spatial, social, and environmental features of built environments, and they are being applied to tasks like optimizing seating in sports venues \cite{Wu_Wang_2024}. In our prior work \cite{Doctorarastoo2024NeurIPS}, we illustrated the use of GNNs to capture human preferences, thereby enriching Reinforcement Learning 
based behavior models. Concurrently, Convolutional Neural Networks (CNNs) with their ability to learn from grid-like spatial inputs and have also been employed for seat recommendations \cite{Moins_etal_2020}. Simpler Multilayer Perceptrons (MLPs) often serve as a baseline, processing features independently or with limited local context, and have been explored in studies identifying factors affecting student seat selection \cite{LosonczyMarshall_2013}. What remains unclear is 
how effectively these different architectural biases translate into models that can predict human spatial preferences not just in existing settings but, more importantly, in new and unseen environmental layouts. Each of these architectures possesses distinct inductive biases regarding how spatial context and features are learned. Establishing which types of architectures are most suitable for achieving scenario-based generalizability is essential for developing adaptable preference models for CPSIS.
This paper addresses this gap by conducting a comparative study of GNNs, CNNs, and MLPs, for human spatial preference prediction. 
This study utilizes synthetic data generated using rule-based agent-based simulation for a case study involving four distinct pocket park layouts. To assess scenario-based generalizability, the leave-one-out cross-validation (LOOCV) strategy is employed. The principal contributions of this work are: (1) a qualitative and quantitative assessment of GNN, CNN and MLP generalizability to unseen environments; and (2) the derivation of insights to guide the selection of appropriate modeling techniques for developing real-world CPSIS.



\vspace{24pt}
\noindent \uppercase{\textbf{METHODOLOGY}} 
\vspace{12pt}

The methodology detailed herein focuses on the comparative evaluation of generalizability of GNN, CNN, and MLP architectures in predicting human spatial preferences to unseen scenarios. 
The core components include the definition of the pocket park environment, data representation, specifications of each preference model, and the experimental protocol for assessing performance. 

\vspace{12pt}
\noindent \textbf{Pocket Park Environment and Data Representation} 
\vspace{12pt}

\begin{figure}[h!]
\centering 

\begin{subfigure}[b]{0.3\linewidth}
    \centering
    \includegraphics[width=\linewidth]{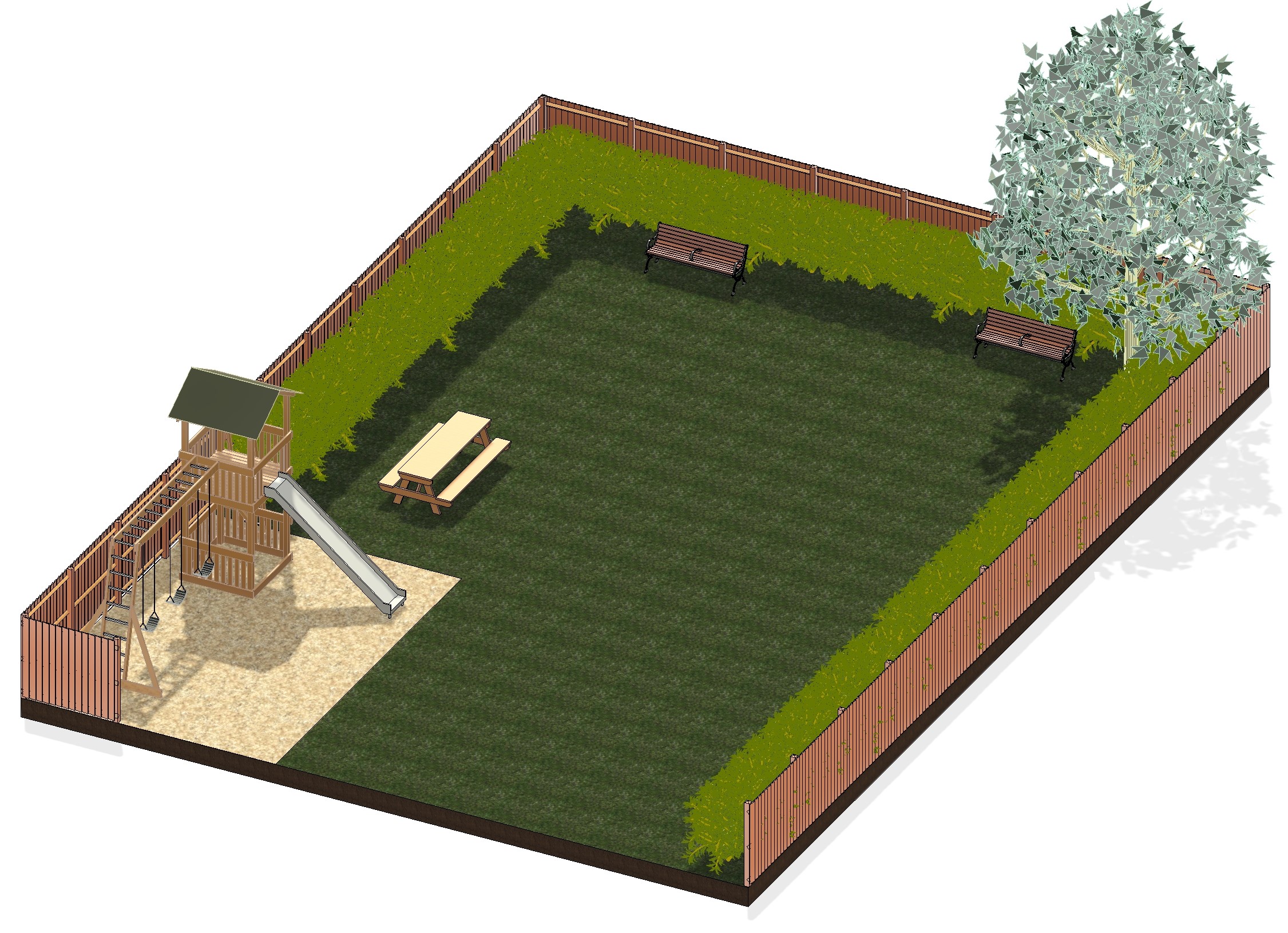}
    \caption{Layout 1}
    \label{fig:layout_a} 
\end{subfigure}
\hspace{0.04\linewidth} 
\begin{subfigure}[b]{0.3\linewidth}
    \centering
    \includegraphics[width=\linewidth]{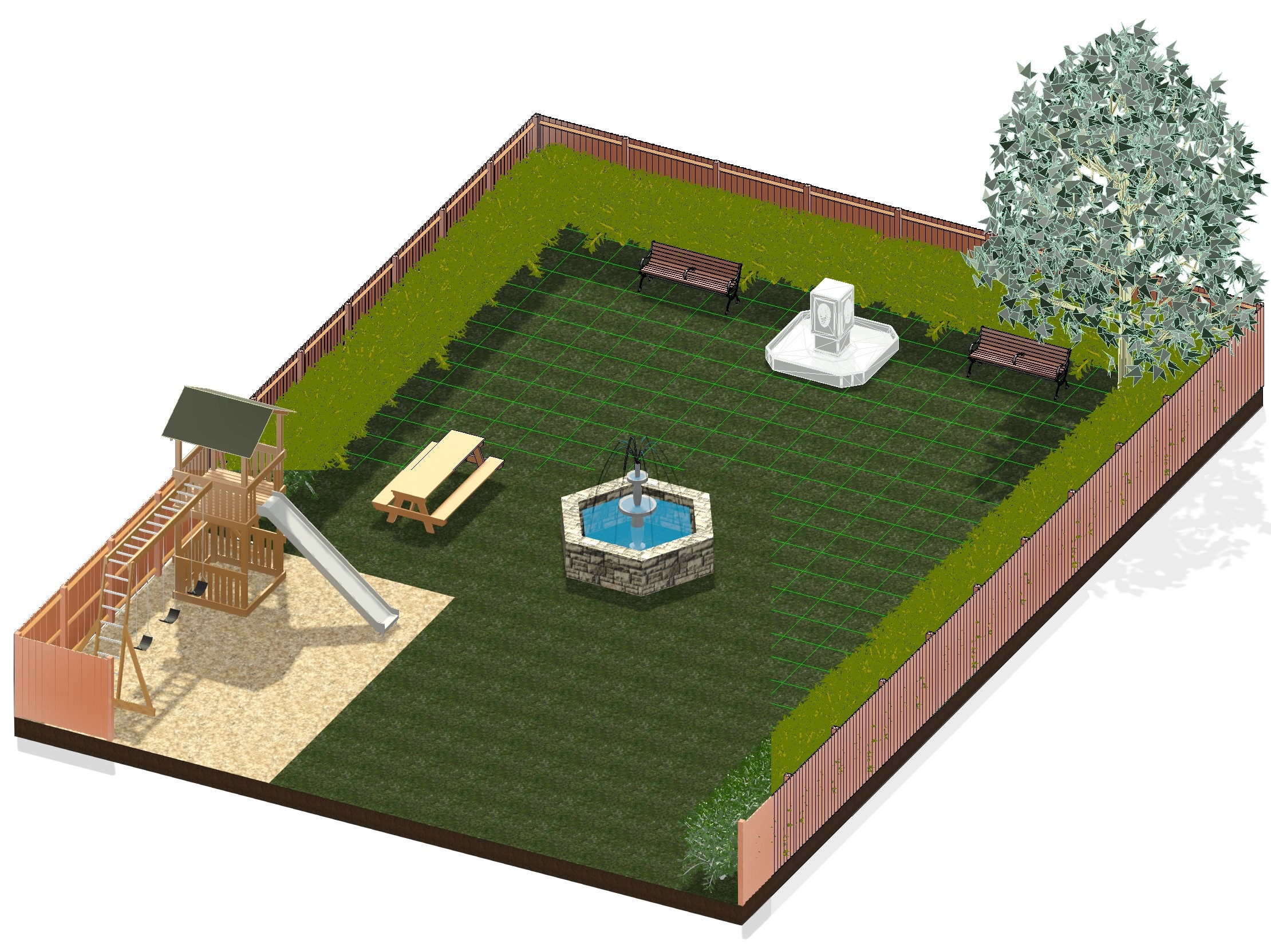}
    \caption{Layout 2}
    \label{fig:layout_b}
\end{subfigure}

\vspace{-0.2em} 

\begin{subfigure}[b]{0.3\linewidth}
    \centering
    \includegraphics[width=\linewidth]{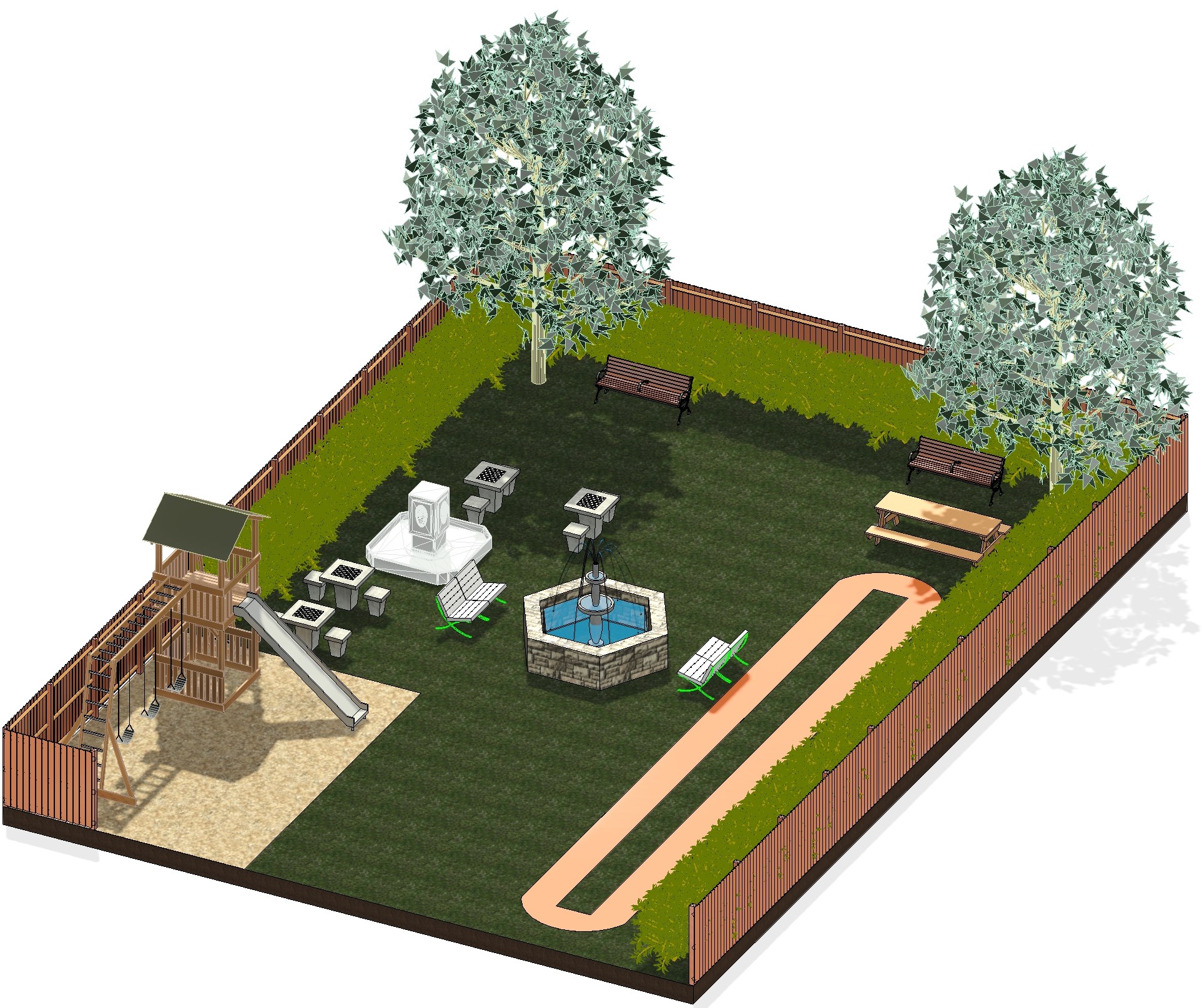}
    \caption{Layout 3}
    \label{fig:layout_c}
\end{subfigure}
\hspace{0.04\linewidth} 
\begin{subfigure}[b]{0.3\linewidth}
    \centering
    \includegraphics[width=\linewidth]{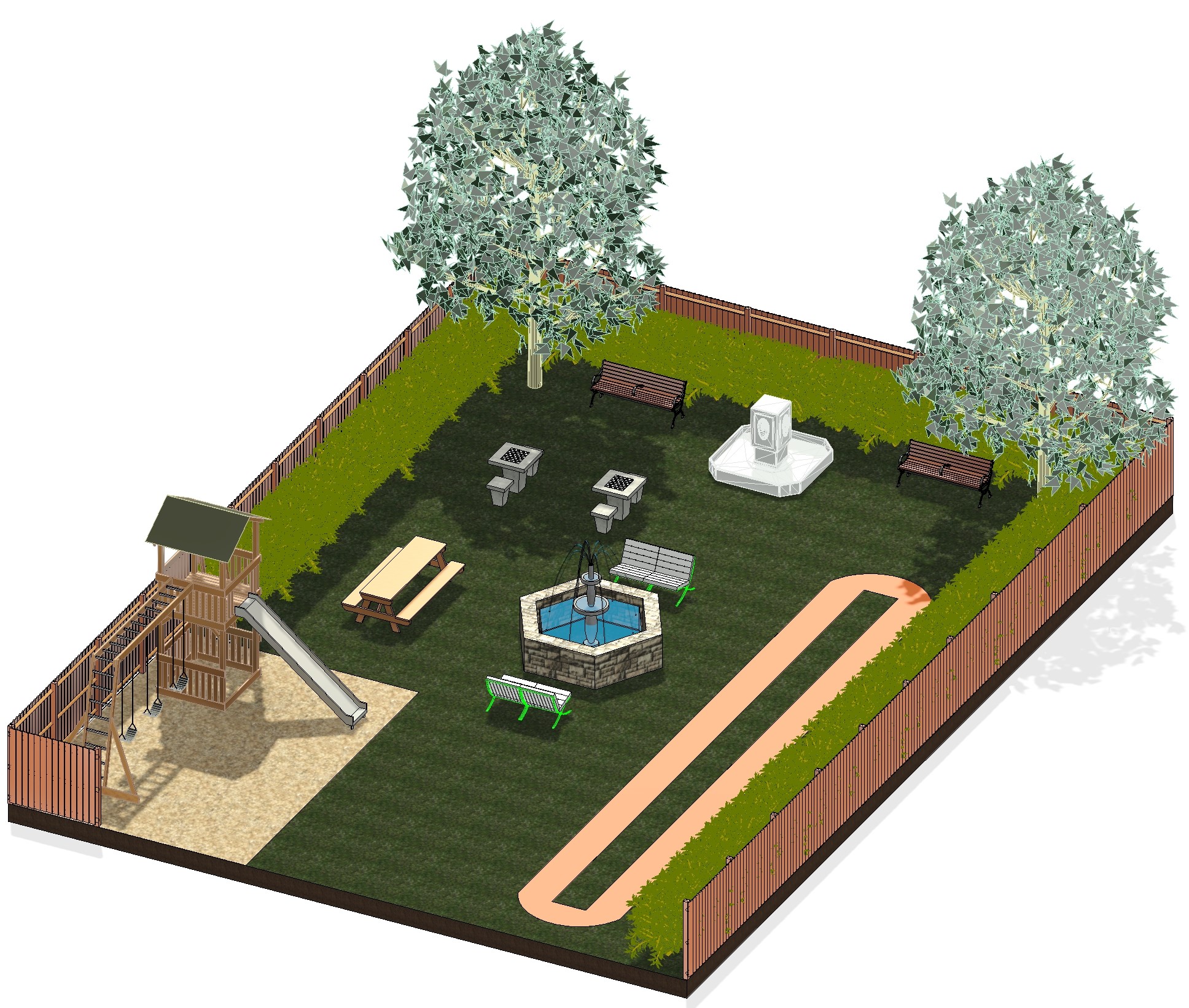}
    \caption{Layout 4}
    \label{fig:layout_d}
\end{subfigure}

\caption{\small Four park layouts used to evaluate generalizability across different scenarios.}
\label{fig:park_layouts}
\end{figure}

\begin{figure}[h]
    \centering \vspace{-0.1cm}
    \includegraphics[width=0.76\linewidth]{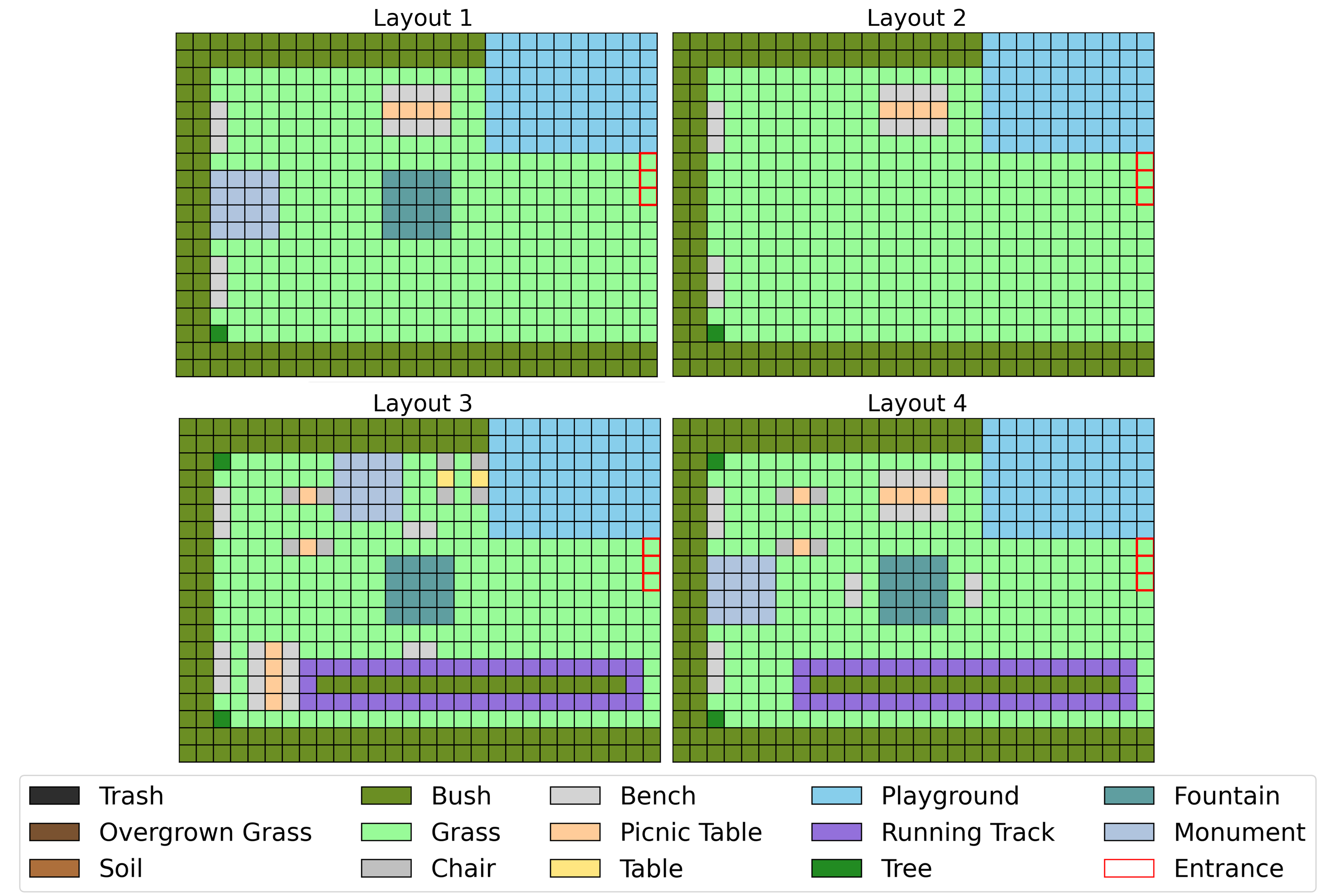}
    \caption{\small Grid-based representations of the park layouts.}
    \label{fig:grid_layouts} \vspace{-0.4cm}
\end{figure}

A simulated park environment is employed as the context of the illustrative case study. We construct four distinct 
layouts (Figure \ref{fig:park_layouts}), designed to test the models' prediction capacity in varying configurations. The park 
measures 15 m $\times$ 21 m and is discretized into 0.75 m $\times$ 0.75 m cells, each representing a potential location for 
activities such as walking on the trail, sitting, eating, or playing in playground (Figure \ref{fig:grid_layouts}).

Each grid cell 
is characterized by a feature vector $x_i$ that encapsulates three categories of attributes. \textit{Physical Features} 
detail the presence and type of park elements such as benches, picnic tables, playgrounds, 
monuments and other amenities; terrain types like grass, soil patches, and running tracks; and obstacles including 
bushes, and trees. \textit{Environmental Features} 
consist of dynamic attributes such as temperature, light intensity, and shadow coverage, which are updated based on simulated time-of-day, solar position, and occlusions from physical structures. Lastly, \textit{Social Features} 
encompass information concerning the presence of other simulated agents within the park. 
\textit{The objective of the preference models is to predict a probability indicating the likelihood of each cell being selected for a given activity, based on the interplay between agent-specific preferences (implicitly learned) and the cell's features.}

\vspace{12pt}
\noindent \textbf{Synthetic Preference Data Generation} 
\vspace{12pt}

Synthetic ground-truth data is generated for this preliminary study due to the unavailability of real-world preference datasets for these specific park layouts. At the same time, using synthetic data allows for full control over the environment and enables a systematic exploration of model behavior and generalizability under varying spatial conditions. A rule-based agent-based simulation algorithm based on by \cite{cheliotis2020agent} is conducted where agents navigate the distinct park layouts, select locations, and engage in activities. These decisions are governed by predefined preference rules that interact with the dynamic physical, environmental, and social attributes of each layout. For training the preference models, the location selected by an agent for a particular activity within these simulations serves as the positive target sample. The input data for the GNN, CNN, and MLP models comprises the feature vector 
for every grid cell, while the target 
is a binary label indicating whether that cell was chosen for the activity. The input and output in training data are intentionally selected to be easily replaced by sensor-collected data with no data from internal human states that are not possible to collect using privacy-preserving sensor employed in the environment.
To increase the diversity of training scenarios, data augmentation is applied to the training samples. This includes 180-degree rotations and planar (horizontal and vertical) flips of the entire park layout grid. 

\vspace{12pt}
\noindent \textbf{Preference Model Architectures}
\vspace{12pt}

To ensure a fair comparison that emphasizes architectural biases, four neural network architectures—a GNN, a 2D CNN (CNN2D), a 1D CNN (CNN1D), and a MLP—were constructed with similar parameter counts (approximately 74k–76k). All models minimize a weighted Binary Cross-Entropy (BCE) loss, suitable for the inherently imbalanced preference prediction task, as only one cell per instance is chosen for any activity. The specific architectural details for each model are summarized in Table \ref{tab:model_architectures}.



\begin{table}[t]\small
\centering
\caption{\small Summary of neural network architectures.}

\label{tab:model_architectures_relative_width}
\begin{tabularx}{\textwidth}{@{} l >{\raggedright\arraybackslash\hsize=0.9\hsize}X >{\raggedright\arraybackslash\hsize=1.1\hsize}X @{}}
\toprule
\textbf{Model} & \textbf{Input Representation} & \textbf{Core Architecture} \\
\midrule
\vspace{-5pt}
GNN & Graph (nodes $=$ cells, edges $=$ 8-way neighbors) & 5-layer GCN \& ReLU, plus a final GCN output layer with Sigmoid \\
\addlinespace
\vspace{-5pt}

CNN2D & 2D grid of the layout with features as channels & 3-layer Conv2D \& ReLU, plus a final 1x1 convolution with Sigmoid \\
\addlinespace
\vspace{-5pt}
CNN1D & Sequence of per-cell feature vectors & 3-layer Conv1D \& ReLU, followed by a Sigmoid activation \\
\addlinespace

MLP & Per-cell feature vector (augmented with 3x3 context) & 4-layer fully-connected network \& ReLU, terminating in a Sigmoid output \\
\bottomrule
\end{tabularx}
\label{tab:model_architectures}
\end{table}

\vspace{12pt}
\noindent \textbf{Experimental Design and Evaluation Metrics} \vspace{12pt}

A LOOCV strategy is central to this study. The process involves four experimental folds. In each fold, data generated from three of the four distinct park layouts are aggregated to form the training set for the GNN, CNN2D, CNN1D and MLP preference models; the data from the single remaining park layout is reserved as the unseen test set for that fold. This rotation ensures that each layout serves as the unseen test scenario once, providing an assessment of generalizability across different spatial configurations.

\textit{Performance Metrics:} The models' performance is assessed using several key metrics. The Weighted BCE Loss 
tracks prediction error during training. For evaluating discriminative ability on this highly imbalanced preference prediction task (only one cell is \textit{chosen} out of many), the Area Under the ROC Curve (ROC AUC) is commonly used. However, ROC AUC can be misleadingly optimistic in scenarios with a large skew between positive and negative classes, as a high number of true negatives (correctly identifying \textit{non-chosen} cells) can increase the score even if the model performs poorly on the rare positive class (\textit{chosen} cells). Therefore, we place a stronger emphasis on the Area Under the Precision-Recall Curve (AUPRC). AUPRC provides a more informative assessment of performance on the positive class, which is of primary interest here, as its baseline is the fraction of positives in the dataset \cite{sofaer2019area}. To quantify the retention of performance when models are applied to unseen layouts, we define Generalizability Scores (GS) per Eq. \ref{eq:GS-AUPRC}, where a GS value closer to 1 signifies superior generalizability of the model to unseen environment layouts:
\beq
GS_{AUPRC} = \frac{\text{AUPRC on Unseen Test Layout}}{\text{Average AUPRC on Seen Training Layouts (validation splits)}}
\label{eq:GS-AUPRC}
\eeq

\vspace{24pt}
\noindent \uppercase{\textbf{RESULTS AND DISCUSSION}} \vspace{12pt} 

This section presents the comparative evaluation of the GNN, CNN2D, CNN1D, and MLP models. We first summarize the final generalization scores to provide a high-level comparison of the architectures, and then analysis of performance variance in each unseen layout. 
All models were trained for up to 100 epochs with a learning rate of $10^{-3}$, employing early stopping with a 30-epoch patience threshold to prevent overfitting. Convergence was monitored using weighted BCE loss on both training data and a held-out validation subset composed of the three seen layouts per LOOCV fold. 


\vspace{12pt}
\noindent \textbf{Comparative Performance on Seen and Unseen Park Layouts} 
\vspace{12pt}

\begin{figure}[b]
    \centering
    \includegraphics[width=0.69\linewidth]{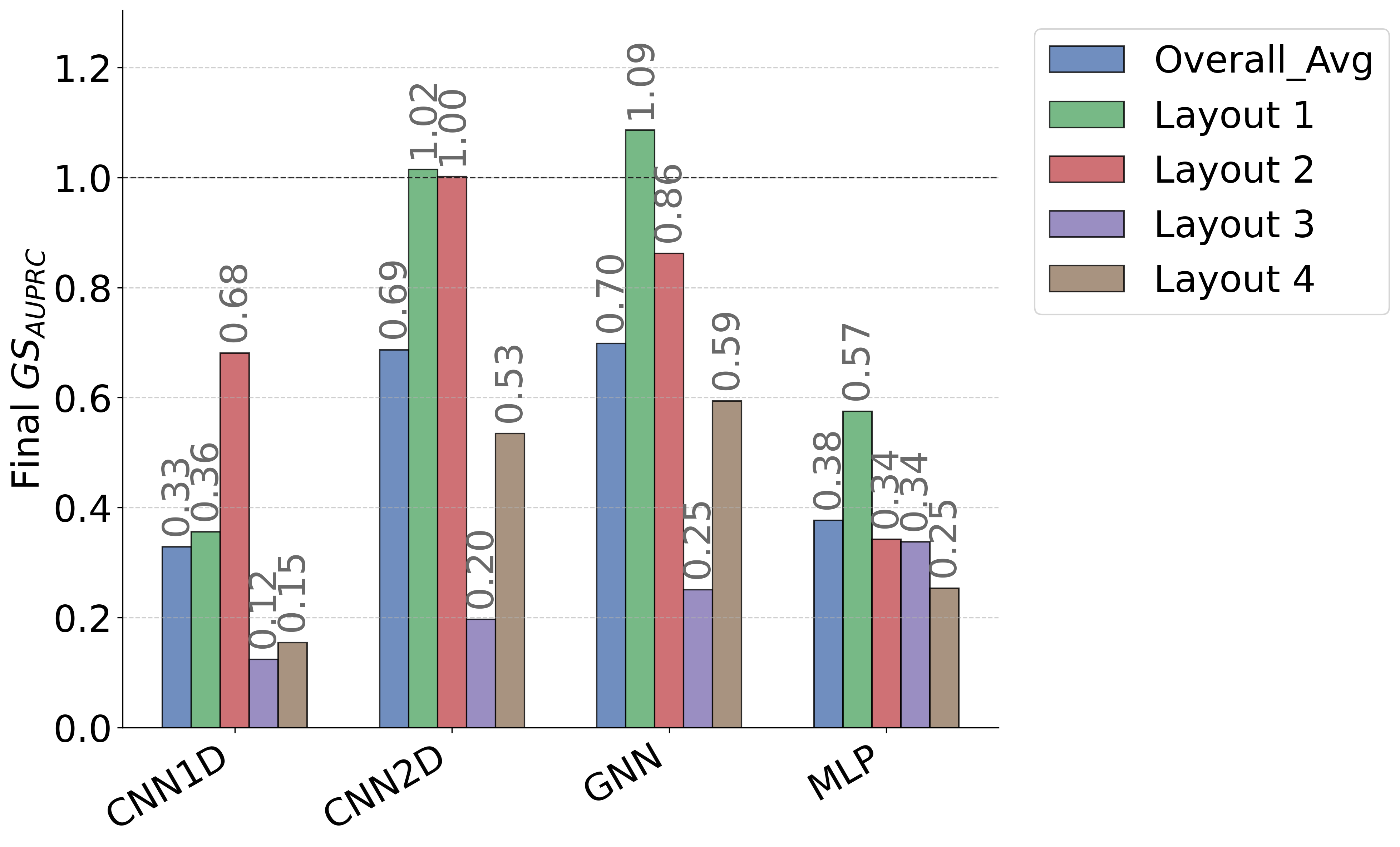}\vspace{-0.25cm}
    \caption{\small Final generalizability score. (GS\textsubscript{AUPRC})} 
\label{fig:gs_barchart}
\end{figure}

The aim of this study is to assess the generalization capability of each architecture to park layouts not encountered during training. Figure~\ref{fig:gs_barchart} presents the GS\textsubscript{AUPRC} averaged across three agents with various preferences for each unseen test layout in the LOOCV protocol, along with the mean over all folds (\textit{Overall\_Avg}). 
The results indicate a clear performance hierarchy. The GNN architecture achieves the highest overall generalization score (Overall\_Avg GS\textsubscript{AUPRC} $\approx$ 0.70), closely followed by the CNN2D (Overall\_Avg GS\textsubscript{AUPRC} $\approx$ 0.69), suggesting both models retain their predictive performance well in unseen layouts. Conversely, the MLP and CNN1D exhibit significantly lower overall scores (0.38 and 0.33, respectively), highlighting their limitations in adapting to new spatial configurations.
Notably, performance varies significantly across the individual layouts. Both GNN and CNN2D demonstrate exceptional generalization on \textit{Layout 1} (GS\textsubscript{AUPRC} $=$ 1.09 and 1.02, respectively) and \textit{Layout 2} (GS\textsubscript{AUPRC} $=$ 0.86 and 1.00, respectively), with scores near or above 1.0 indicating that their performance on unseen tests was on par with or exceeded their performance on the validation data. In contrast, all architectures struggled with \textit{Layout 3}, which consistently yielded the lowest GS\textsubscript{AUPRC} across all models (all $\leq$ 0.25). The denser arrangement of \textit{Layout 3} may have introduced unfamiliar spatial patterns, contributing to reduced generalization performance.

\begin{figure*}[t!]
\centering
\begin{subfigure}[b]{0.48\textwidth}
    \centering
    \includegraphics[width=\textwidth]{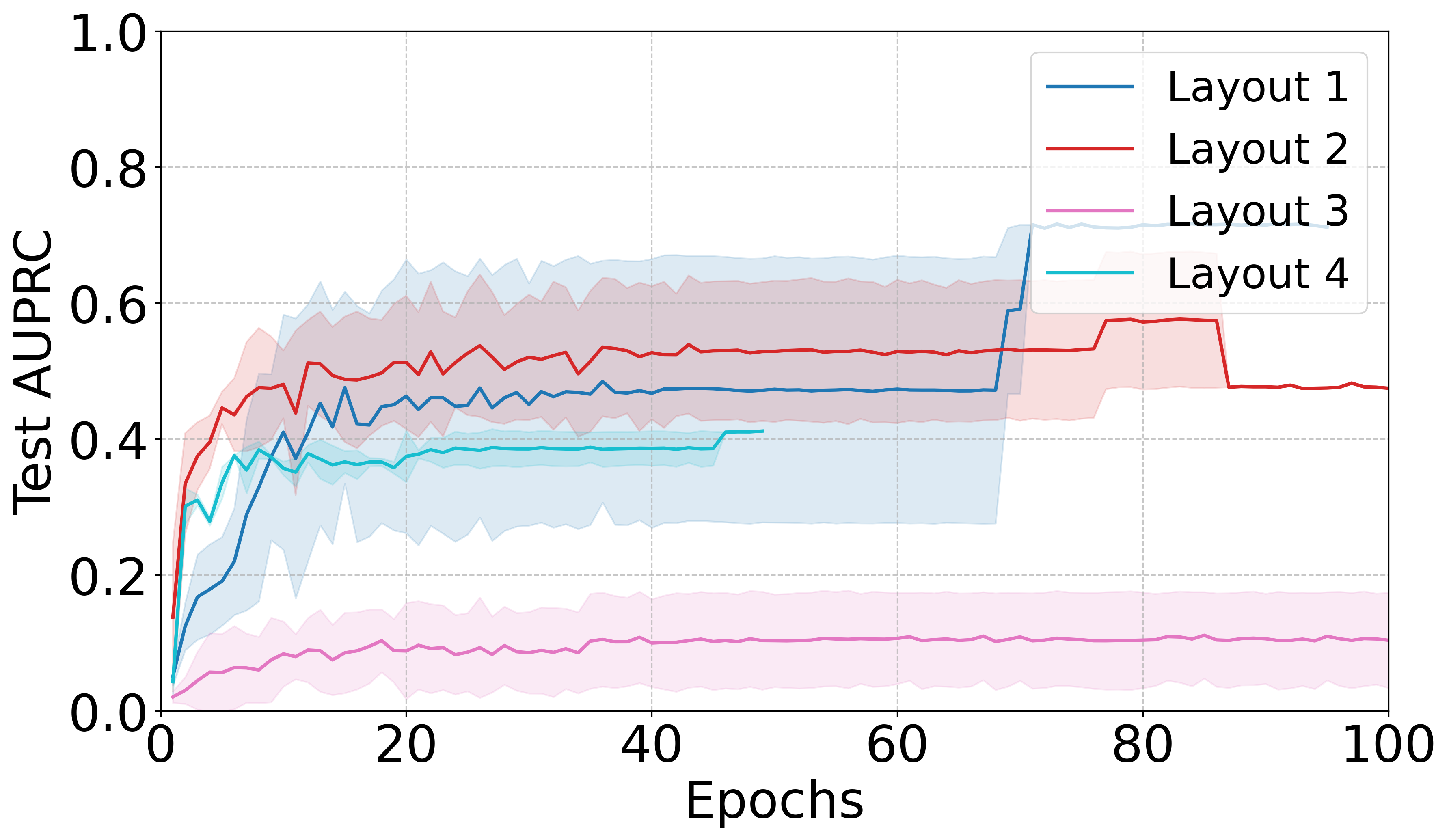}\vspace{-0.25cm}
    \caption{GNN Performance by Layout}
    \label{fig:gnn_by_env}
\end{subfigure}
\hfill
\begin{subfigure}[b]{0.48\textwidth}
    \centering
    \includegraphics[width=\textwidth]{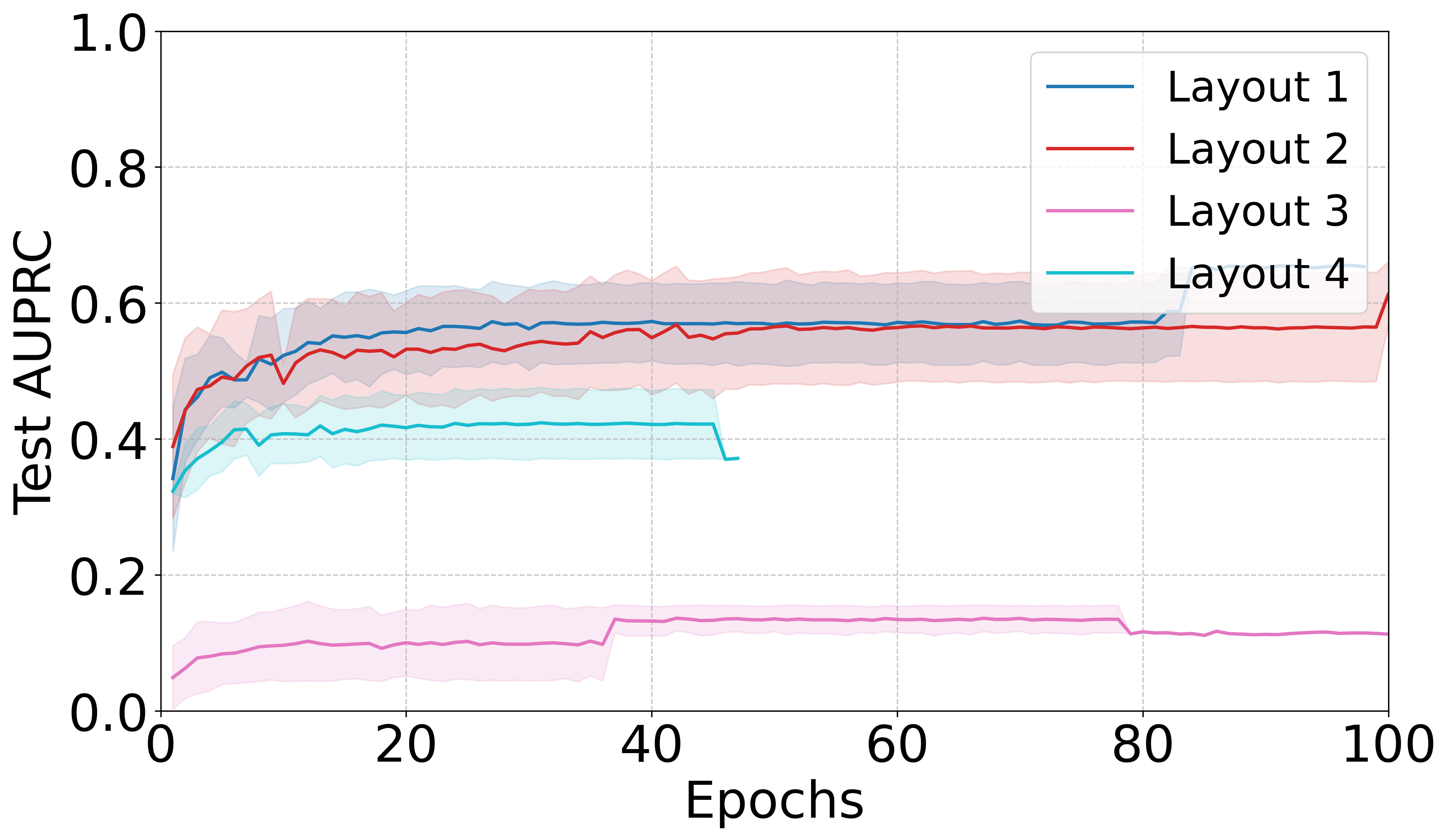}\vspace{-0.25cm}
    \caption{CNN2D Performance by Layout}
    \label{fig:cnn2d_by_env}
\end{subfigure}


\begin{subfigure}[b]{0.48\textwidth}
    \centering
    \includegraphics[width=\textwidth]{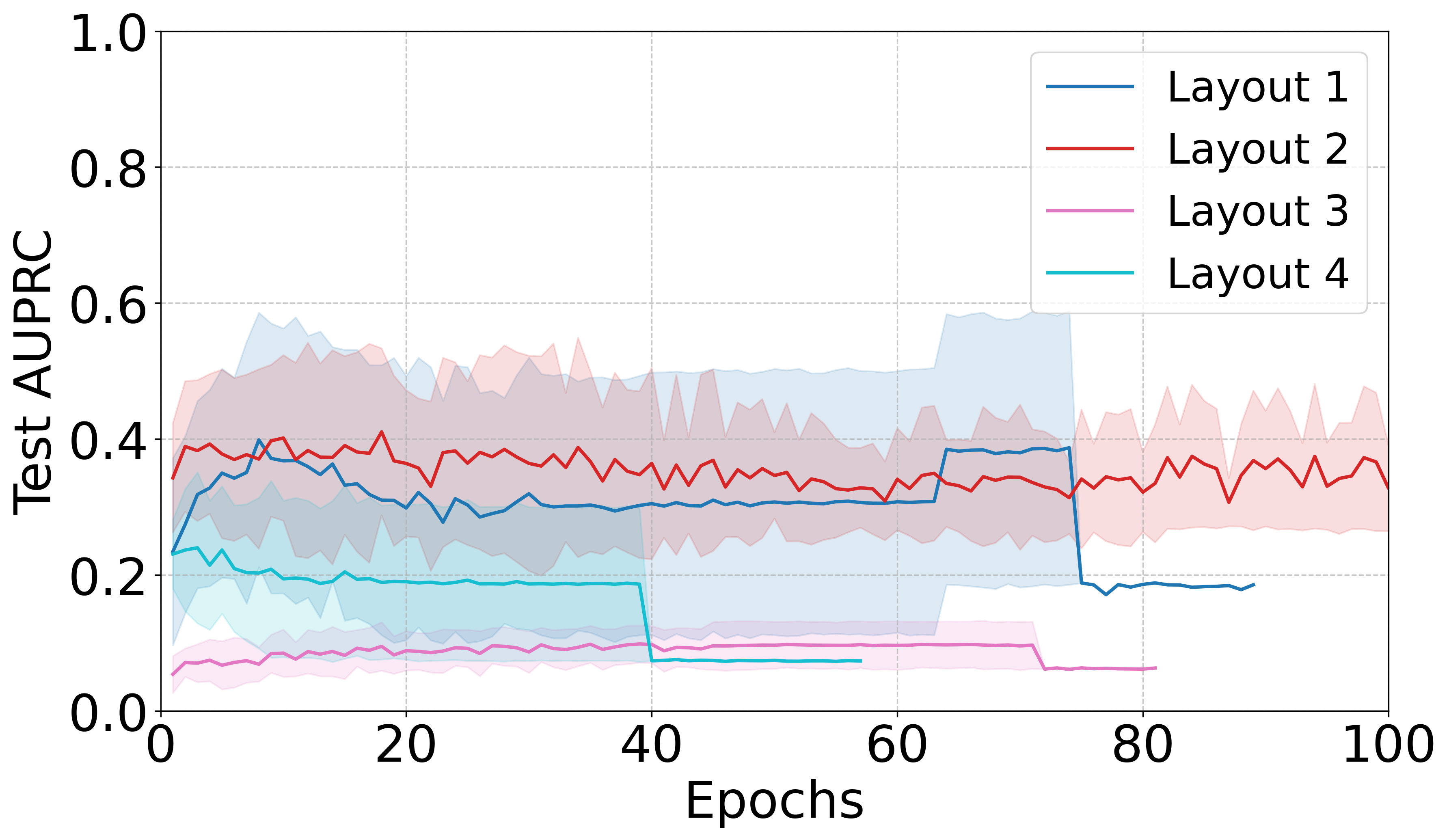}\vspace{-0.25cm}
    \caption{CNN1D Performance by Layout}
    \label{fig:cnn1d_by_env}
\end{subfigure}
\hfill
\begin{subfigure}[b]{0.48\textwidth}
    \centering
    \includegraphics[width=\textwidth]{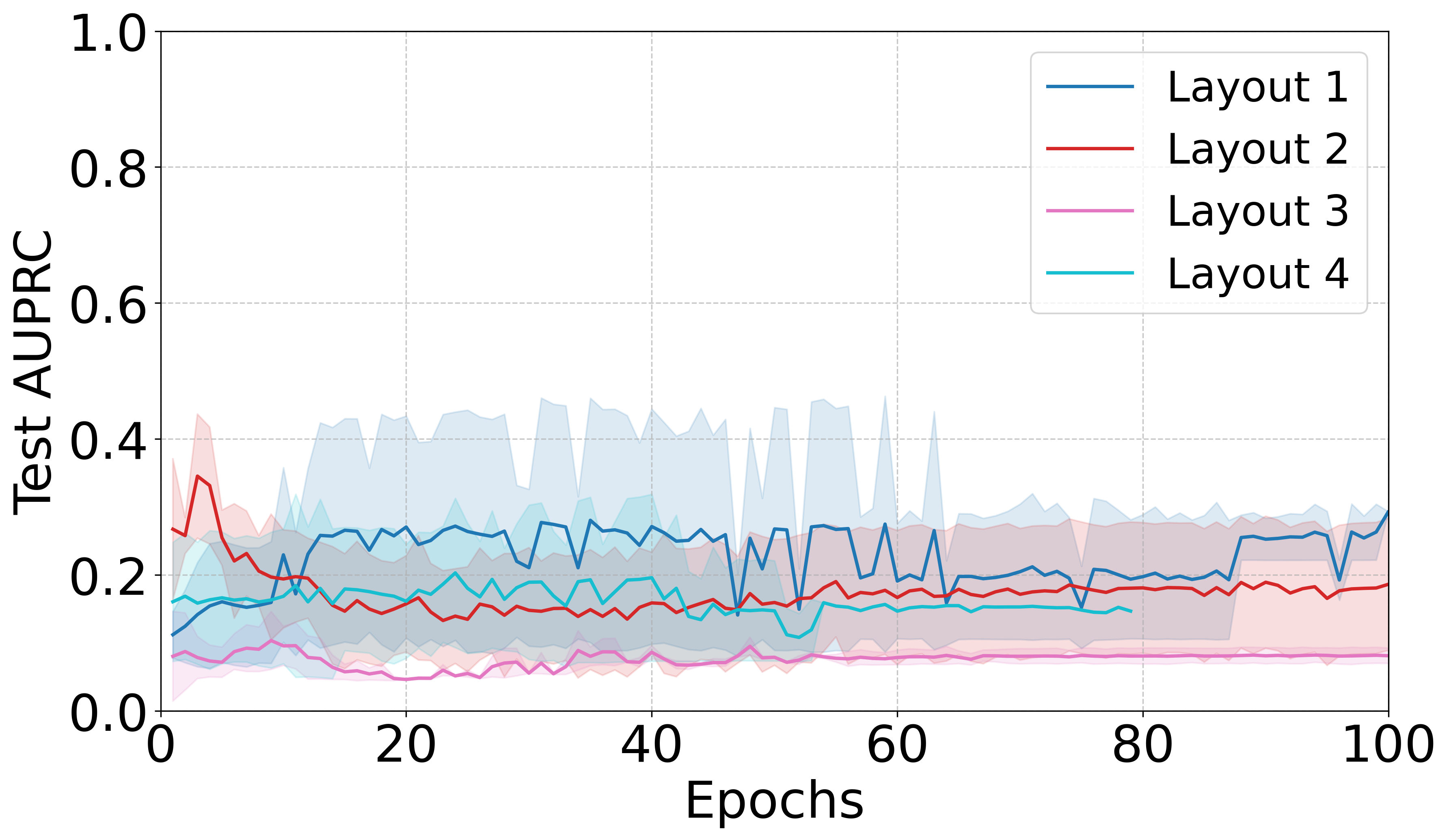}\vspace{-0.25cm}
    \caption{MLP Performance by Layout}
    \label{fig:mlp_by_env}
\end{subfigure}
\caption{\small Average Test AUPRC vs. Epochs for each model architecture. The shaded regions represent the standard deviation across agents.}
\label{fig:per_model_performance}
\end{figure*}

To further analyze model behavior, Figure~\ref{fig:per_model_performance} illustrates the average Test AUPRC over the training epochs for each model architecture, with each curve showing the model's performance on one of the four layouts when it served as the unseen test set. Models with strong spatial inductive biases, GNN and CNN2D (Figures~\ref{fig:per_model_performance}a and~\ref{fig:per_model_performance}b, respectively), exhibit high and consistent AUPRC scores (approximately 0.5–0.6) on \textit{Layouts 1} and \textit{2}, moderate performance on \textit{Layout 4}, and correctly identify \textit{Layout 3} as particularly challenging (AUPRC~$<$~0.2). These trends suggest that these models have learned meaningful and generalizable spatial features. In contrast, the CNN1D and MLP models (Figures \ref{fig:per_model_performance}c and \ref{fig:per_model_performance}d) display highly unstable and noisy performance across all layouts, failing to converge to a strong solution. The MLP, in particular, shows very little variance between the different layouts; it performs equally poorly on all of them, with all curves clustered at a low AUPRC ($<$0.3). This behavior reinforces the conclusion that while models with strong spatial priors can generalize to diverse, unseen scenarios, to different extents, the simpler models lack the architectural robustness required for this task.


\vspace{24pt}

\noindent \uppercase{\textbf{CONCLUDING REMARKS}} \vspace{12pt}

This study underscores the importance of architectural inductive biases in achieving scenario-based generalizability for spatial preference prediction. GNN and CNN2D models' ability to maintain predictive performance in unseen layouts highlights their promise for real-world CPSIS applications, where environments vary dynamically. By leveraging synthetic yet controlled environments, this work lays the groundwork for evaluating and deploying preference-aware models that can adapt to spatial diversity—a critical step toward closing the loop between human behavior and infrastructure design.

\vspace{24pt}
\noindent \uppercase{\textbf{Acknowledgments}} 

\vspace{12pt}

This work is support by the National Science Foundation under Grant \#2425121.

\vspace{24pt}

\small 

\bibliographystyle{iwshm}
\bibliography{IWSHM_references}


\end{document}